\documentclass[final]{cimento}
\title{Photoexcitation of the Nucleon in the Constituent 
      Quark Model}
\author{I.~P.~Matamba\from{ins:ven}, M.~Braun\from{ins:unisa} 
\atque S. A. Sofianos\from{ins:unisa}}
\instlist{\inst{ins:ven} 
Physics Department, University of Venda, Private Bag X5050,
Thohoyandou 0950, South Africa
\inst{ins:unisa}
Physics Department, University of South Africa,
P.O.Box 392, Pretoria 0003, South Africa 
}
\begin{document}
\maketitle
\begin{abstract}
Photoexcitation of the nucleon is studied in the framework of the 
constituent quark model with nonrelativistic quark-quark 
interactions. Utilizing the dominance of the pairwise structure of the
relevant three-quark  potential, the bound and excited state wave functions 
have been constructed  using the Faddeev-type Integrodifferential Equation 
Approach. The method is based on the  Hyperspherical Harmonics expansion and  
takes two-body correlations exactly into account and therefore reliable 
wave functions, required in photo processes, can be obtained.
The  integrated photoabsorption cross sections
for the $E1$ and $M1$ transitions are calculated and  compared with 
experimental values as well as with those of other theoretical calculations. 
Our results, in overall, are in good to very  good agreement with the 
experimental values.
\PACSes{
\PACSit{12.39}{Quark models} 
\PACSit{12.39.Pn}{Potential models}
\PACSit{14.20.Gk}{Baryon resonances}
}
\end{abstract}
%%%%%%%%%%%%%%%%%%%%%%%%%%%%%%%%%%%%%%%%%%
\section{Introduction}
In the nonrelativistic constituent quark model for mesons and baryons the  
quarks are bound by confining potentials. Despite their limitations and
concerns about their validity, these potential models describe the
properties of the various mesons and baryons surprisingly well (see, for 
example, Silvestre-Brac \cite{Brac1} and the  two excellent review articles 
by Lucha et al. \cite{Lucha} and Richard \cite{Richard} on the matter). 

Once the potential model approach is adopted, the three-quark problem  
can be solved using various approaches. Among them, the Hyperspherical 
Harmonics (HH) method is quite successful in applications as it is well
suited to describe  radial and  orbital excitations 
\cite{FabreL,Bada1,Bada2,Haysak}. Within the framework of the HH method, 
the Hypercentral Approximation (HCA) has been used in the past 
\cite{FabreF,McTav,VHothers} to study the spectrum of the baryon.
There are various reasons for adopting the HCA
to study the three quark system:  
i) The two-body potential acting between quarks is quite soft and 
therefore in the HH expansion of the interaction only 
the first term gives a significant contribution to the binding energy. This 
of course means that the two-body correlations are not as strong as 
compared to the nuclear correlations; 
ii) it is quite simple and thus one avoids the complicated three-body
calculations via, for example, the Faddeev equations \cite{Brac1,Fad}, and
iii) the results obtained from it are accurate and the spectra are well 
reproduced.

Another  method,  in the framework of the HH method,  is the   
Integrodifferential Equation Approach (IDEA) \cite{Fmodels,FFS,Oe1,Oe2}
which includes  higher terms  of the HH  expansion in an average manner.  
The IDEA  method  takes  two-body correlations into account exactly, 
reproduces the spectrum of the nucleon quite well, and provides 
wave functions reliably \cite{Oe1,Oe2} which is  crucial in studying 
photoexcitation processes. These  processes are manifested as  resonances and
can be excited through electromagnetic transitions giving rise to large 
enhancements in the total absorption cross section \cite{MMG}.

The photoexcitation of the nucleon resonances has been studied in the past 
by various groups \cite{MMG,Brizol,Isgur}. The results obtained by them are 
rather unsatisfactory when compared to the experimental data. The inclusion 
of retardation effects and relativistic corrections does not improve the
situation much  \cite{MMG,Brizol}. In this work we consider the absorption 
of a single photon by a nucleon which then undergoes a transition from the 
ground state to an excited one.  The  photoabsorption cross section is 
calculated using various quark-quark potentials and by using the HCA and 
IDEA methods.

In Sec. 2  we describe our formalism. In Sec. 3  we give details 
on how the $E1$ and $M1$ transition amplitudes are calculated
while in Sec. 4  we present our results  and discussions.

\section{Formalism}
%%%%%%%%%%%%%%%%%%%%
The  photoexcitation process is described by the transition amplitude
\begin{equation}
      M_{fi}=<\Psi_f|H|\Psi_i>\,,
\label{trans}
\end{equation}
where  $\Psi_i$  is the initial ground  state wave function of the nucleon,
 $\Psi_f$ is the wave function of  the final excited state,  and $H$ 
the perturbative electromagnetic Hamiltonian. In what follows we shall 
discuss these ingredients in some detail.

\subsection{The wave functions}
The fully antisymmetric total wave function for a  three-quark system 
can be expressed  as a product of configuration  space, flavor, spin, and color
functions. Since baryons are color singlets,  the  color wave  
function is totally antisymmetric ($A$) and thus the remaining product must be
fully symmetric ($S$),
\begin{equation}
     \Psi_{\mbox{total}}^A= \underbrace{
	\psi_{\mbox{space}}\times \Phi_{\mbox{flavor}} \times
       \chi_{\mbox{spin}}}_S
       \times \underbrace{C_{\mbox{color}}}_A\,.
\label{Psi}
\end{equation}
The structure of the  symmetric component of Eq. ({\ref{Psi}) depends
on the transition considered and can be constructed using the 
various symmetries involved.\\

For the construction of the symmetric part of the total wave function the 
fully symmetric,  mixed symmetric, and mixed antisymmetric configuration 
space wave functions are required. These can be obtained using
the IDEA \cite{Fmodels,FFS} method. In this  method the fully  symmetric 
ground state configuration space wave function is constructed 
from the Faddeev-type components $P(z,r)$ \cite{Oe2} 
\begin{equation}
    \Psi^S (\vec{\rho},\vec{\sigma}) = \frac{1}{r^{5/2}} \left [
	   P^S(z_{12},r)+ P^S(z_{23},r) + P^S(z_{31},r) \right ]\,,
\label{IDEAS}
\end{equation}
where ($\vec{\rho},\vec{\sigma}$) are  the Jacobi coordinates,    
$
 r=\left [\frac{2}{3} \sum_{\alpha} \rho_{\alpha}^2 
\right]^{1/2} 
$
is the hyperradius with $\rho_\alpha = r_\alpha\,\,, \alpha=12,23,31$,   and 
the  $z_{\alpha}$ are given by
$
	z_{\alpha} = 2\rho_{\alpha}^2/r^2 -1\,.
\label{za}
$
The required mixed symmetry states for $L=1$  are given by
\begin{eqnarray}
	 \Psi_1^{M^S}(\vec{\rho},\vec{\sigma}) & = & \frac{1}
		       {r^{5/2}} \bigg \{(1+z_{12})^{1/2}Y_{10}(\omega_{12})
         P_1^{S^\prime}(z_{12},r) \nonumber \\ 
    & - & \frac{1}{2} \bigg [(1+z_{23})^{1/2}Y_{10}(\omega_{23})
           P_1^{S^\prime}(z_{23},r)  \\
     & + & (1+z_{31})^{1/2}Y_{10}(\omega_{31})
          P_1^{S^\prime}(z_{31},r) \bigg ] \bigg \} \,,\nonumber \\
	 \Psi_1^{M^A}(\vec{\rho},\vec{\sigma}) & = & \frac{1}{r^{5/2}} 
    \bigg [(1+z_{31})^{1/2}Y_{10}(\omega_{31})P_1^{S^\prime}(z_{31},r) 
        \nonumber \\
   & - & (1+z_{23})^{1/2}Y_{10}(\omega_{23})P_1^{S^\prime}(z_{23},r) 
     \bigg ] \,,
\end{eqnarray}
where the superscripts $M^S$ and $M^A$ denote the mixed symmetric and
antisymmetric states with respect to the interchange of particles 1 and 2.

The required symmetric spin-flavor states are given by
\begin{equation}
       \left |\xi^S \right > = \frac{1}{\sqrt{2}} 
                         \left [ \Phi^{M^S} \chi^{M^S}
			+ \Phi^{M^A} \chi^{M^A} \right ]\,,
\end{equation}
while the mixed symmetric  states are
\begin{eqnarray}
 \left |\xi^{M^S} \right > & = & \frac{1}{\sqrt{2}} 
	   \left [ \Phi^{M^S}\chi^{M^S}- \Phi^{M^A} \chi^{M^A} \right ]\,,\\
 \left |\xi^{M^A} \right > &=& \frac{1}{\sqrt{2}} 
	   \left [ \Phi^{M^S} \chi^{M^A} + \Phi^{M^A} \chi^{M^S} \right ]\,.
\end{eqnarray}

The relevant flavor and spin states are given by various authors and 
therefore, will not be presented  here (see, for example, Refs. 
\cite{Isgur,CLOSE}).    
The singlet, antisymmetric color state,
\begin{equation}
      C_{color}^A = \frac{1}{\sqrt{6}}(RBY-BRY+BYR-YBR+YRB-RYB)\,,
\end{equation}
where R,B and Y stand for Red, Blue, and Yellow respectively, does not enter 
into the calculations and therefore, in what follows will be suppressed. 

The initial total wave function for the proton (P)
ground state, with $L=0,\, S=1/2$, and $J=1/2$,  is given by
\begin{equation}
       \left|\Psi_i\right>  =  \frac{1}{\sqrt{2}}\left [
	    \Phi^{M^S}_{\rm P} \chi^{M^S}_{\rm P}+\Phi^{M^A}_{\rm P} \chi^{M^A}_{\rm P} \right]
       \left|\Psi^S_{0}\right>\,,
\label{PROTONG}
\end{equation}
where  the lower index of the space wave function $|\Psi_0^S>$  refers to 
the angular momentum $L$. The final wave function for the first excited state,
with $L=1, \,
S=1/2$, and $J=1/2 \,\, \mbox{or} \,\,  3/2$, of the proton is
\begin{equation}
 \qquad  \big  |\Psi_f \big >  =  \frac{1}{2}\bigg [
     \bigg ( \Phi^{M^S}_{\rm P} \chi^{M^S}_{\rm P}-\Phi^{M^A}_{\rm P}
     \chi^{M^A}_{\rm P}\bigg) \big |\Psi^{M^S}_{1}  \big >
+  \bigg ( \Phi^{M^S}_{\rm P} \chi^{M^A}_{\rm P}+\Phi^{M^A}_{\rm P}
     \chi^{M^S}_{\rm P}\bigg)\big|
     \Psi^{M^A}_{1}\big > \bigg ]\,.
\end{equation}
For the $M1$ transition $ (S=1/2) \rightarrow
(S=3/2)$, where the proton and $\Delta^+(1232)$
both have an angular momentum  $L=0$, the total wave function for the initial 
state of the proton is still given by Eq. (\ref{PROTONG}), while  the final wave function
for the delta is
\begin{equation}
      \left|\Psi_f \left>=\Phi_\Delta^S
      \chi_\Delta^S\right |\Psi_{\Delta}^S\right>\,.
      \label{FWP}
\end{equation}
%
%%%%%%%%%%%%%%%%%%%%%%%%%%%%%%%%%%%%%%%%%%%%%%
\subsection{Electromagnetic transitions}
The perturbative Hamiltonian  for the electric dipole E1 transition,  in 
the case of three quarks of equal mass m, is given by 
\begin{equation}
	   H_{E1} =  - \frac{1}{mc} \sum_{j=1}^3 \lambda_j
	    \hat{\epsilon}_{\gamma} \cdot \vec{p}_j\,,
\label{HE11}
\end{equation}
where  $ \vec{p}_j$ is the momentum of quark $j$ and $\hat{\epsilon}_{\gamma}$ 
denotes a polarization direction of the incident photon. For $u$ and $d$ quarks, 
the charge operator has the form
\begin{equation}
	    \lambda_j = \frac{e}{6}(1+3\tau_j^z)\,,
\end{equation}
where $\tau_j^z $ is the third component of the isospin of the $j$-th quark
with
\begin{equation}
       	\tau_j^z|u> = |u> ,\qquad \tau_j^z|d> = -|d>\,.
\end{equation}
Using commutation relations to express the  momenta in terms of the 
three-quark Hamiltonian \cite{Roy} we may rewrite (\ref{HE11}) in a Siegert 
form 
\cite{Siegert,Ellerkmann}
\begin{equation}
   H_{E1} =  - \frac{i}{\hbar c}(E_f-E_i)\sum_{j=1}^3 \lambda_j
            \hat{\epsilon}_{\gamma} \cdot \vec{x}_j\,,
\end{equation}
where  $\vec{x}_j$ is the coordinate of $j$-th quark conjugate to the momentum
$\vec{p}_j$.
In Jacobi coordinates we have
\begin{equation}
 H_{E1}=-\frac{1}{2i\hbar c}(E_f-E_i)(\hat{\epsilon}_{\gamma}\cdot\vec{
   \rho})I_p - \frac{\sqrt{3}}{3i\hbar c}(E_f-E_i)(\hat{\epsilon}_{\gamma}
    \cdot\vec{\sigma})I_q\,,
    \label{JC1}
\end{equation}
where $(E_f-E_i)$ is the difference between the final and initial binding 
energies   and  the operators $I{\rm p}$ and $I_q$ are given by  
\begin{eqnarray}
      I_p & = & \frac{e}{2}(\tau_1^z - \tau_2^z), \nonumber \\
      I_q & = & \frac{e}{2}\left(\frac{\tau_1^z + \tau_2^z}{2} -
		\tau_3^z\right)\,.
\end{eqnarray}
Thus, instead of expressing the Hamiltonian in terms of the individual 
particle charge and coordinates, the more appropriate Jacobi coordinates 
and operators $I_p$ and $I_q$ which act on quasi-particles \cite{Sandhas}
are used.

The magnetic dipole $M_1$ causes the transition       
\hspace{1mm} $\gamma {\rm P} \rightarrow \Delta^+(1232)$,\hspace{1mm}
in which  the proton, after absorbing a photon ($\gamma$), is excited to
the delta ($\Delta^+$). The corresponding perturbative Hamiltonian is
\begin{equation}
      H_{M1}=-i\sum^3_{j=1}\left(\vec{\mu}_q^j\times \vec{k_\gamma} \right)
        \cdot \hat{\epsilon}_{\gamma}\,,
\end{equation}
where  $\vec{\mu}_q^j$ is the magnetic moment operator of the $j$-th quark.
Since $H_{M1}$ does not contain any orbital operators, in this transition
the spin must change instead.
%%%%%%%%%%%%%%%%%%%%%%%%%%%%%%%%%%%%
\section{ The transition amplitudes}
Noting that $E_\gamma = E_f - E_i$ and by letting the charge operators $I_p$ 
and $I_q$ act on the three-quark isospin states we finally obtain for the 
transition matrix elements
\begin{equation}
     M_{E1}  =  \frac{eE_\gamma}{2\sqrt{6}i\hbar c}\bigg(\left<\Psi^{M^A}_{1}
      \left |\hat{\epsilon}_\gamma \cdot \vec{\rho} 
      \right|\Psi^S_{0}\right> - \left < 
      \Psi^{M^S}_{1}\left |\hat{\epsilon}_\gamma \cdot
      \vec{\sigma}\right|\Psi^S_{0}\right >\bigg)\,.
\label{ME11}
\end{equation}
The integrals in (\ref{ME11}) were  evaluated using Euler angles  
 $\alpha,\,\beta,\,\gamma$  as external and $\rho,\,\sigma,\, x= 
{\vec{\rho}\cdot\vec{\sigma}}/{\rho \sigma}$  as internal coordinates. 
Here the  $z^\prime$  axis is chosen to coincide with $\hat{\rho}$ and 
$\hat{\sigma}$ is lying in the $x^\prime - z^\prime$  plane. Thus only a
five dimensional integration has to be done numerically since both $\Psi_0^S$ 
and the two components of  $\Psi_1$ are invariant with respect to a rotation 
about the z-axis and thus do \it not} depend on $\alpha$. 
After averaging over the direction of $\vec{k}$ and polarization direction
one obtains the following expression for the absolute square of the 
transition matrix elements
\begin{equation}
	\overline{\big|{\cal M}_{E1}\big|^2} = \frac{e^2E_\gamma^2}
	{72\,(\hbar c)^2}\bigg|\left<\Psi_1^{M^A}\big|\rho_z 
	\big|\Psi_0^S\right> - \left<\Psi_1^{M^S}
	\big|\sigma_z \big|\Psi_0^S\right>\bigg |^2\,. 
\end{equation}
Therefore, the following integrals are required
\begin{eqnarray} 
	& & \left<\Psi_1^{M^A}\big|\rho_z \big| \Psi_0^S \right>  
	=  2\pi \int_0^\pi \sin \beta d\beta  \int_0^{2\pi}d\gamma 
	\int_0^\infty \rho^2 d\rho \int_0^\infty \sigma^2 d\sigma 
	\int_{-1}^1 dx \nonumber \\ 
 	& & \quad \qquad \qquad \qquad \Psi_1^{M^A}(\rho,\sigma,x,\beta,\gamma)
	\rho \cos \beta     \, \Psi_0^S(\rho,\sigma,x)\,, \\
    	&  &  \nonumber    \\
	& & \left<\Psi_1^{M^S}\big|\sigma_z \big| \Psi_0^S \right> =  
	2\pi \int_0^\pi \sin \beta d\beta\int_0^{2\pi}d\gamma \int_0^\infty 
	\rho^2 d\rho \int_0^\infty \sigma^2 d\sigma
	\int_{-1}^1 dx \nonumber \\ 
	& &\qquad \Psi_1^{M^S}(\rho,\sigma,x,\beta,\gamma) \sigma (\cos 
	\beta \cos \theta -\sin \beta \cos \gamma \sin \theta )\,\Psi_0^S
	(\rho,\sigma,x)\,. 
\end{eqnarray}
The corresponding  integrated photoabsorption cross section
for a single excited electric dipole
state, in the long wavelength limit \cite{Brizol,Eisen}, is
\begin{equation}
	\Sigma_1 = \int dE_\gamma\sigma_\gamma^{E1}=\frac{4\pi^2\hbar c}
    	{\hbar\omega}\,\overline{\big |{\cal M}_{E1} \big |^2} \,.
\end{equation}
For the $M1$ transition amplitude we have

\begin{equation}
	 M_{M1}  =  <\Psi_f|H_{M1}|\Psi_i>
       =  i\left(\hat{\epsilon}_\gamma \times \vec{k_\gamma} \right)
	 \sum^3_{j=1}<\Psi_f| \mu_q^j \sigma^z_j |\Psi_i>\,,
\end{equation}
where  $\Psi_i$ and $\Psi_f$ are given by Eqs. (\ref{PROTONG}) and (\ref{FWP})  
respectively. 

The process  \hspace{0.7mm} $\gamma P \rightarrow \Delta^+(1232)$
($\frac {1}{2}^+ \rightarrow \frac {3}{2}^+ $) \hspace{0.3mm}
can take place  either via  the magnetic dipole ($M1$) or  the
electric quadrupole ($E2$) transition. In the quark model the latter transition is
forbidden \cite{Becchi} because it is proportional to the charge
operator which cannot cause transitions  between quark spin $1/2$ and $3/2$
states, and hence the matrix element vanishes by orthogonality of the quark
spin wave functions.  The $M1$ transition involves the quark magnetic moments -- hence
the spin operator -- and this can lead to transitions $ (S=1/2) \rightarrow
(S=3/2)$ \cite{Dalitz}. The transition matrix element can be written as
\begin{equation}
    M_{M1}=i\left(\hat{\epsilon}_\gamma \times \vec{k_\gamma} \right)
   <\Psi_f|\mu_q^1 \sigma^z_1 + \mu_q^2 \sigma^z_2 + \mu_q^3 \sigma^z_3
   |\Psi_i>\,.
\end{equation}
Using the the flavor, spin, and configuration space wave functions 
and  averaging over the two photon polarization directions we finally obtain

\begin{equation}
	\overline{ \big|{\cal M}_{M1}\big|^2}=\frac{2\alpha}{9}\frac{E^2_\gamma 
	\hbar c}{(mc^2)^2}I^2_{M1} \,,
\end{equation}
where $I_{M1}$ is the overlap integral given by
\begin{equation}
	I_{M1}=8\pi^2 \int_0^\infty\rho^2d\rho\int_0^\infty\sigma^2d\sigma
	\int_{-1}^1 dx\, \Psi_\Delta^S(\rho,\sigma,x)\Psi_0^S(\rho,\sigma,x)\,.
\end{equation}
Like in the electric transitions, the photoabsorption cross
section for a single excited magnetic dipole state is
\begin{equation}
    	\Sigma_{M1}= \int dE_\gamma\sigma_\Delta^{M1}=\frac{4\pi^2\hbar c}
    	{\hbar\omega}\,\overline{\big|{\cal M}_{M1} \big|^2} \,.
\end{equation}
%
%%%%%%%%%%%%%%%%%%%%%%%%%%%%%%%%%%%%%%%%%%%%%%%%%%%%
%%%%%%%%%%%%%%%%%%%%
\section{Results and Discussions}
%%%%%%%%%%%%%%%%%%%%%%%%%%%%%%%%%%%%%%%%%%%%%%%
The  quark-quark  potential can be written as a sum of the central
and the spin-spin parts
\begin{equation}
		V_{qq} = V^c + V^s\,,
\end{equation}
where $V^c$ contains, as usual, the confinement and the coulombic parts while 
 $V^s$ is of the general form 
\begin{equation}  
		V^s = f_{ij}(r)\vec{\sigma}_i\cdot\vec{\sigma}_j\,.
\end{equation}
{ The spin-dependent interaction, suggested by the perturbative one-gluon 
exchange mechanism, is important in describing  the splitting of 
the meson masses  and the experimental  mass difference
$M_\Delta - M_N$.  It contains the function $f_{ij}(r)$ which is singular 
as $\delta(r)$ and therefore the corresponding wave equation has no physically 
acceptable solutions and one expects a collapse in both the quark-quark and in 
baryon systems. To avoid this a cut-off or a smearing function is introduced to
reduce the singularity and to make the calculations tractable. 
Since the resulting  spin-spin potential is strongly attractive and
short-ranged, it can generate significant short--range correlations which
have  stronger effects  on the ground state  than on the excited $ L > 0 $ 
states. The strong  correlations at short distances in the $L=0$ 
partial wave are not expected to play a major role in  photoexcitation processes
as the excited states  are shifted to the outer region and thus in the overlap 
integral short-range  contributions  are rather  unimportant. Therefore  
any influence of the spin-spin force will come from the modification
of the mass differences and of the corresponding wave functions.  
}
In this work we performed calculations with the $V^c$ term alone as 
well as with both terms included. For the calculations with the $V^c$  
part only,  we employed the Martin \cite{Martin,Richard2}, the Cornell
\cite{eichten,Bada3}, and the Lichtenberg \cite{Licht} potentials.
For the calculations in which  both  terms are included we used the
Ono-Sch\"{o}berl potential \cite{Ono} and
two of the recently published  potentials by Silvestre-Brac \cite{Brac1},
namely, the AP1 and AP2 versions which have the general form
\begin{eqnarray}
	V_{q\bar q}(r_{ij})&=&-\frac{\kappa(1-\exp(-r_{ij}/r_c))}
		{r_{ij}}+\lambda  r_{ij}^p-\Lambda \nonumber\\
         &+ & \frac{2\pi}{3m_im_j}\kappa^{\prime}(1-\exp(-r_{ij}/r_c))
        \frac{\exp(-r_{ij}^2/r^2_0)}{\pi^{3/2}r^3_0}\vec{\sigma}_i\vec{\sigma}_j
\label{AP2}
\end{eqnarray}
with 
$$
	r_0(m_i,m_j) = A\left(\frac{2m_im_j}{m_i + m_j}\right)^{-B}.
$$
The parameters for AP1 are given by
$$
\begin{array}{llll}
	& p = 2/3\,,  &  r_c=0\,,  &  m_u=m_d=0.277\,{\rm GeV}\,,\\
	& \kappa = 0.4242\,, &  \kappa^{\prime}=1.8025\,, &
	\lambda = 0.3898\,{\rm GeV}^{5/3}\,,\\
	& \Lambda = 1.1313\, {\rm GeV}\,, &  B=0.3263\,, &
	A=1.5296\,{\rm GeV}^{B-1}\,, \end{array}
$$
while those of AP2 are given by
$$
\begin{array}{llll}
	& p= 2/3\,, &  r_c=0.3466\,{\rm GeV}^{-1}\,,  &  m_u=m_d=0.280\,
	{\rm GeV}\,,\\ & \kappa = 0.5743\,, & \kappa^{\prime}=1.8993\,,
	&  \lambda = 0.3978\,{\rm GeV}^{5/3}\,,\\
	& \Lambda = 1.1146\, {\rm GeV}\,, &  B=0.3478\,, & A=1.5321\,
	{\rm GeV}^{B-1}\,. 
\end{array}
$$
We remind here the reader that  the quark-quark potential  $V_{qq}$ is  
related to the quark-antiquark potential $V_{q\bar q}$ by Lipkin's rule
\cite{Green}, i.e., $ V_{qq} = V_{q\bar{q}}/2$.

The results obtained for the ground state and the first orbitally excited 
state of the nucleon  with  spin-independent potentials and by using the
IDEA and HCA methods are given in table I. Both methods are in very  good 
agreement with each other and the transition energy $E_1-E_0$, { shown
in table II,}  is reasonably well reproduced. This transition  energy and 
the corresponding wave functions  were used  to calculate the integrated 
photoabsorption cross section $\Sigma_1$  due to the E1 transition, from 
the ground state to the  N(1520) resonance.  Our results are given in table 
II together with experimental values and  those of other methods. It is seen 
that for both the IDEA and the HCA  the  cross sections  are in good to very 
good agreement with the experimental data. 

{  The results obtained by other methods for the transition energy are  
generally very low as compared to the experimental ones except those obtained 
via the Isgur-Karl (I.K.) model \cite{Isgur}, which, nevertheless, can not  reproduce the experimental 
photoabsorption cross section $\Sigma_1$ well.  This mainly implies that the 
corresponding wave functions are not adequate to describe the photoabsorption 
cross section for the E1 transition. The latter is also true for the wave
functions employed by Brizzolara and Giannini in their thorough investigations 
on the nucleon photoabsorption \cite{Brizol} based on the nonrelativistic 
quark model of Isgur and Karl and the bag model.
They explicitly demonstrated 
that the results are rather strongly  dependent  on the hyperfine mixing and 
on the dimensions of the system and thus on the h.o parameter used. 
The discrepancy with the experimental results still persists even when other 
contributions such as charge plus  current densities with and without retardation 
effects are incorporated into the h.o model. Three-body force components of the 
quark interaction  (in the 3q model), and relativistic corrections in the 
framework of the bag model  do not improve matters to a significant degree
either. Apart, of course, from the inadequacy of the wave functions these 
discrepancies probably stem also from some  other fundamental absorption 
mechanism, such as the coupling to mesons and/or $q\bar q$ pairs \cite{Brizol}.
 }

The  results  for the nucleon masses obtained by using  spin-dependent potentials 
are given in table III. The corresponding transition energy $E_1-E_0$  
together with the  cross sections are presented in table IV. It is seen that
both AP1 and AP2  potentials of Silvestre-Brac give excellent results for the
photoabsorption cross section  $\Sigma_1$.  This came as a surprise since the 
transition energies are not as good as one would like them to be. This can be 
attributed to the fact that in the confining potential the excited state lies
at a higher energy as compared to the experimental value and thus  the wave 
function for $L=1$ is more spread out in space and therefore the overlap
integral somehow compensates  for the excess transition energy.
Our  results obtained with the Ono-Sch\"oberl potential are in fair
agreement with the experimental data.

The delta masses are given in table V while the transition energy $E_\Delta
- E_N$ is shown in table VI; it is  well reproduced but the  results for the
corresponding integrated  photoabsorption cross section of the
$\Delta$(1232) resonance  are  only in fair agreement with the experimental
values. The same is true for the I.K model. This discrepancy  can be connected  
to the strong dependence on the quark masses that enter the cross section 
via the magnetic moments.

{ Finally the energy and cross--section results  obtained via the IDEA and 
HCA methods  are quite similar. This comes as no major surprise as
the quark-quark potentials are soft and thus  two-body 
correlations do not play a  significant role. This means that the corresponding 
wave functions are fairly similar and the small differences between them 
do not strongly influence  the overlap integral and  thus the cross sections.
In the case where a spin-spin force was used the short--range correlations are
not manifested in the overlap integral either due to the shifting of the $L>0$ 
wave function to the outer region.}

In conclusion the nonrelativistic potential model reproduced
the experimental data for the E1 transitions  quite successfully.
There is a weak potential dependence and in the case of the 
spin-independent potentials  the best results are  obtained with the
Lichtenberg potential. In the case of the spin-dependent potentials both
AP1 and AP2 give excellent results for the E1 photoabsorption cross sections.
In the M1 transitions where there is a strong dependence on the quark masses 
the Ono-Sch\"oberl potential gives better results than the AP1 and AP2 potentials.

\solong
Financial support from the Foundation for Research Development 
is  gratefully acknowledged.

\newpage
%%%%%%%%%%%%%%%%%%%%%%%%%%%%%%%%%%%%%%%%%%%%%%%%%%%%%%%%%%%%%%%%%%%
%%%%%%%%%%%%%%%%%%    REFERENCES    %%%%%%%%%%%%%%%%%%%%%%%%%%%%%%%
%%%%%%%%%%%%%%%%%%%%%%%%%%%%%%%%%%%%%%%%%%%%%%%%%%%%%%%%%%%%%%%%%%%

%%%%%%%%%%%%%%%%%%%%%%%%%%%%%%%%%%%%%%%%%%%%%%
%%%%%%%%%%%%%%%%%  TABLES %%%%%%%%%%%%%%%%%%%%
%%%%%%%%%%%%%%%%%%%%%%%%%%%%%%%%%%%%%%%%%%%%%%
\newpage

%Table 1
\begin{table}
\caption { Nucleon masses in MeV obtained with  spin-independent potentials.}
\begin{narrowtabular}{1.2cm}{cccccccccc} \hline
&  Quantum & & &   & IDEA & & & HCA &   \\ 
&  numbers & & Expt& & Potentials & & &Potentials &   \\
L & S & $J^P$  & values & M & C & L & M & C
& L \\ \hline
0 & 1/2 & ${\frac{1}{2}}^+$  & 940 & 1085 & 1087 & 1085 & 1090 & 1090 &
1088 \\
1 & 1/2 & ${\frac{3}{2}}^-$  & 1520 & 1676 & 1773 & 1730 & 1676 & 1774 &
1730 \\ \hline
\end{narrowtabular}
\vspace{5mm}
M : Martin potential; C : Cornell potential; L : Lichtenberg potential. 
\end{table}

%Table 2
\begin{table}
\caption { The integrated photoabsorption
cross section $\Sigma_1$ for the N(1520)
proton resonance  obtained with spin-independent potentials. 
Abbreviations as in table I, see also text.}
\begin{narrowtabular}{2mm}{cccccccc} \hline
 & &  & IDEA &  & & HCA &  \\ 
& Expt & &  Potentials & & &  Potentials &  \\
 & values & M & C & L &  M & C & L  \\ \hline
$E_1-E_0$ & 580 & 591 & 686 & 645 &  586 & 684 & 642  \\
(MeV) &  &  &  &  &  &  &   \\
$\Sigma_1$ & $33^a$ & 30.4 & 27.8 & 30.6  & 30.6 & 28.6 & 31.3  \\ 
(MeVmb) &  &  &  &  & &  &     \\ \hline
& & & OTHER & METHODS  & & &  \\ \hline
 &  & I.K. & h.o. & h.o. & h.o. & 3q & bag   \\
 &  &  &  & (ret) & ($\rho+j$) &  &   \\ \hline
$E_1-E_0$ & 580  & $595^b$ & 167 & 167 & 167 & 144 & 349   \\
(MeV) &  &  &  &  &  &  &   \\
$\Sigma_1$ & $33^a$ & $26^c$ & $40^c$ & 38 & 43 & 27 & 13   \\
(MeVmb) &  &  &  & &  &  &    \\ \hline 
\end{narrowtabular}
$^a$Ref. \cite{Armstrong}; $\quad$
$^b$Ref. \cite{Isgur}; $\quad$
$^c$Ref. \cite{Brizol}.\\
The symbols ret, $\rho$, and j mean  retardation, classical charge density
and  current density respectively.
\end{table}

%Table 3
\begin{table}
\caption { Nucleon masses in MeV obtained with spin-dependent potentials.}
\begin{narrowtabular}{2mm}{cccccccccc} \hline
& Quantum &  &  & & IDEA & & & HCA &  \\ 
& numbers & &Expt   & & Potentials & &  & Potentials & \\
L & S & $J^P$  & values & AP1  & AP2 & OS &  AP1 & AP2 & OS \\ \hline
0 & 1/2 & ${\frac{1}{2}}^+$  & 940 & 1028 & 1021 & 920 & 1035 & 1029 & 940 \\
1 & 1/2 & ${\frac{3}{2}}^-$  & 1520 & 1772 & 1797 & 1821  & 1773 & 1797 & 1821 \\ \hline
\end{narrowtabular}
\vspace{5mm}
OS : Ono-Sch\"oberl potential; AP1, AP2 : Silvestre-Brac potentials.
\end{table}

%Table 4
\begin{table}
\caption { The integrated photoabsorption
cross section  $\Sigma_1$ for the N(1520)
proton  resonance obtained with spin-dependent potentials.
Abbreviations as in table III, see also text.}
\begin{narrowtabular}{2mm}{cccccccc} \hline
 & & & IDEA & &  & HCA & \\ 
& Expt & & Potentials & &  & Potentials & \\
 & values & AP1 & AP2 & OS  & AP1 & AP2 & OS  \\ \hline
$E_1-E_0$ & 580 & 744 & 776 & 901  & 738 & 768 & 881  \\
(MeV) &  &  &  &  &  &  &   \\
$\Sigma_1$ & $33^a$ & 32.6 & 32.3 & 27.1  & 32.9 & 32.5 & 27.4  \\ 
(MeVmb) &  &  &  &  &  &  &   \\ \hline
\end{narrowtabular}
\end{table}

%Table 5
\begin{table}
\caption{Delta masses in MeV obtained with spin-dependent
potentials. Abbreviations as in table III, see also text.}

\begin{narrowtabular}{2mm}{cccccccccc} \hline
 &  Quantum  &  & & & IDEA & & & HCA & \\ 
 & numbers  &  & Expt & & Potentials  & & & Potentials & \\
 L & S & $J^P$  & values & AP1  & AP2 & OS  & AP1 & AP2 & OS \\ \hline
0 & 3/2 & ${\frac{3}{2}}^+$  & 1232 & 1300 & 1307 & 1224  & 1300 & 1308 &
1225 \\ \hline
\end{narrowtabular}
\end{table}

%Table 6
\begin{table}
\caption {The integrated photoabsorption cross section $\Sigma_{M1}$ for
the $\Delta$(1232) resonance obtained with spin-dependent potentials.
Abbreviations as in tables II and III, see also text.}
\begin{narrowtabular}{2mm}{ccccccccc} \hline
 & & & IDEA &  & & HCA & &  OTHER METHODS \\ 
 & Expt & & Potentials  & & &  Potentials  & & \\
 & values & AP1 & AP2 & OS  & AP1 & AP2 & OS  & I.K. \\ \hline
$E_\Delta-E_N$ & 292 & 272 & 286 & 304  & 265 & 279 & 285  & $300^b$ \\
(MeV) & & & &  &  &   &  &   \\
$\Sigma_{M1}$ & $63^a$ & 81 & 83 & 54  & 79 & 82 & 58  & 52 \\ 
(MeVmb) &  &  &  &  &  &    \\ \hline
\end{narrowtabular}
\end{table}
\end{document}